\documentclass{edp-conf}
\usepackage{graphicx}
\def\arcsec{\hbox{$^{\prime\prime}$}}

\def\gsim{~\rlap{$>$}{\lower 1.0ex\hbox{$\sim$}}}
\def\lsim{~\rlap{$<$}{\lower 1.0ex\hbox{$\sim$}}}
\begin{document}

\TitreGlobal{SF2A 2004}

\title{The CLEF-SSH simulation project}
\author{da Silva, A.C.}
\address{IAS, B\^atiment 121, Universi\'e Paris Sud, F-91405 Orsay, France}
\author{Kay, S.T.}
\address{Astronomy Centre, University of Sussex, Falmer, Brighton BN1 9QH, UK}
\author{Aghanim, N.$^1$}
\author{Blanchard, A.}
\address{LATT, Av. Edouard Belin 14, F-31500 Toulouse,France}
\author{Liddle, A.R.$^2$}
\author{Puget, J.-L.$^1$}
\author{Sadat, R.$^3$}
\author{Thomas, P.A.$^2$}
\runningtitle{CLEF-SSH simulation project }
\setcounter{page}{1}
\index{Author1, A.}
\index{Author2, B.}
\index{Author3, C.}

\maketitle
\begin{abstract} 
The CLEF-SSH simulation project is an international collaboration,
involving the IAS, LATT and the University of Sussex, to produce large
hydrodynamic simulations of large-scale structure that implement
realistic models of radiative gas cooling and energy feedback. The
objective is to use these simulations to study the physics of galaxy
clusters and to construct maps of large angular size of the
Sunyaev--Zel'dovich (SZ) effect, for the preparation of future
experiments. Here we present results from a first run, the CLEF
hydrodynamics simulation, which features $2(428)^3$ particles of gas
and dark matter inside a comoving box with 200 $h^{-1}$ Mpc on a side.
\end{abstract}
%

\section{The CLEF-SSH simulation project}

Future cluster surveys probing different wavebands, such as in the
X-ray and the microwave spectral regions (see e.g. Romer et
al. (2001), Carsltrom et al. (2003)), will provide invaluable
information on the study of galaxy clusters and the formation of
cosmological structure (e.g. Viana et al. (2003), Carstrom, Holder and
Reese (2002)). Hydrodynamic $N$-body simulations are undoubtedly one
of the best tools for the preparation and scientific interpretation of
these surveys. They permit a deeper understanding of galaxy cluster
physics and much-improved quality of simulated maps for dedicated
instrument studies.

The CLEF-SSH (CLuster Evolution \& Formation using Supercomputer
Simulations with Hydrodynamics) project is a collaboration between the
Institut d'Astrophysique Spatiale (IAS, France), the Laboratoire
d'Astrophysique Toulouse Tarbes (LATT, France), and the University of
Sussex (Astronomy Centre, UK), to perform large simulations of
large-scale structure (LSS), using state-of-the-art hydrodynamic
techniques and realistic models of cosmological gas physics.  Our aim
is to use these simulations to perform studies of galaxy clusters with
improved statistics and to construct large SZ maps for dedicated
instrument studies.

The project was awarded with 100,000 CPU hours at CINES, the French
national parallel computing centre in Montpellier, during
2004. Sixty-six per cent of this time was consumed to produce a first
run (the CLEF simulation) with $2(428)^3$ particles of gas and dark
matter inside a comoving volume of $(200\,h^{-1}\,{\rm Mpc})^3$.  The
simulation was generated using a parallel version of the code Gadget
II (Springel, Yoshida \& White 2001) and includes radiative gas
cooling and an energy feedback model described in Kay (2004). It
assumes a flat $\Lambda$CDM cosmology with cosmological parameters
matching present observations: matter density $\Omega_{{\rm m}}=0.3$,
baryon density $\Omega_{{\rm B}}=0.0486$, hubble parameter $h=0.7$,
and $\sigma_8=0.9$. The mass of each gas and dark matter particles are
$m_{\rm gas}=1.4\times 10^{9}\, h^{-1}\,M_\odot$ and $m_{\rm
dark}=7.1\times 10^{9}\, h^{-1}\,M_\odot$, respectively.  With these
characteristics (see Kay et al (2004) for further simulation details),
CLEF is one of the largest existing simulations with radiative cooling
and energy feedback. This allowed us to generate extensive cluster
catalogues over a wide range of mass and redshift. Our cluster
catalogue at redshift $z=0$ contains more than a thousand objects with
mass $M\gsim 5\times 10^{13}\,h^{-1}\,M_\odot$, each resolved with at
least 7000 particles. The design of CLEF permits us to address several
important issues of the formation and evolution of cosmological
structure. In the next sections we present preliminary results
regarding cluster scaling relations and simulated maps of the SZ
effect from CLEF.

\section{Cluster scaling relations}

In Kay et al. (2004) we present X-ray cluster scaling relations from
the CLEF simulation at $z=0$. These show an overall good agreement
with observations. Here we concentrate on the correlation between the
cluster integrated SZ flux, $Y = \int y \,\,d\Omega$ ($y$ is the
comptonization parameter) and X-ray luminosity, $L_{\rm X}$. This
relation permits an estimation of the SZ fluxes of known X-ray
clusters (for example in future Planck observations) without the need
to resort to scalings that involve mass determinations.

\begin{figure}[t]
\centering \includegraphics[width=8.8cm]{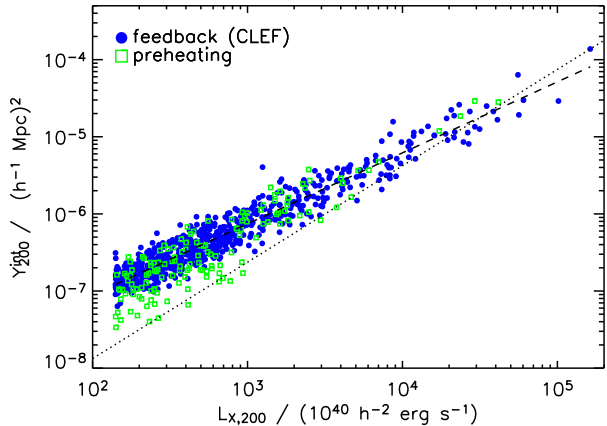}
\caption{Cluster integrated SZ flux versus bolometric X-ray luminosity
of clusters at z=0 in CLEF (dots) and the preheating simulation in da
Silva et al. (2004) (open squares). The dashed line is a power-law best
fit to the CLEF cluster distribution. The dotted line illustrates the
self-similar scaling normalised to the most luminous cluster.}
\label{fig_1}
\end{figure}

In Fig.~\ref{fig_1} we plot the {\it intrinsic} SZ flux, $Y^{\rm
int}=Y\,d_A^2$ ($d_A$ is the angular diameter distance), versus
bolometric X-ray luminosity of clusters from the CLEF run at $z=0$
(filled circles). Quantities are evaluated within spherical regions where
the mean overdensity is 200 times the critical density. We use $Y^{\rm
int}_{200} = \frac{\sigma_{\rm T}\,k_{\rm B}}{m_{\rm
e}c^2}\,\frac{0.88}{m_{\rm p}}\, \sum\limits \, m_i \, T_i$ and
$L_{\rm X, 200} = \sum \, m_i \, \rho_i \,\Lambda(T_i,Z)/(\mu m_{\rm
p})^2$, where $m_i$ and $T_i$, are the mass and temperature of the
particle $i$, summations run over hot ($T_i>10^5$K) gas particles, and
$\Lambda(T_i,Z)$ is the bolometric cooling function, see e.g. da Silva
et al. (2004).
The dashed line is a power-law best fit to the cluster distribution,
\begin{equation}
Y^{\rm
int}_{200}=(6.2\pm 0.1)\times 10^{-6}\,\left( { L_{\rm x,200}/L_{44} }
\right)^{0.92\pm 0.01}\,\left( { h^{-1}\,{\rm Mpc} } \right)^2,
\label{eq1}
\end{equation}
where $L_{44}=10^{44}\, h^{-2}\, {\rm erg\, s^{-1}}$. The slope of
this relation deviates significantly from the slope predicted by the
self-similar model, which assumes a purely gravitational scenario of
cluster formation. To illustrate this deviation, we also plot the
self-similar scaling, $Y^{\rm int}\propto L_{\rm x}^{1.25}$,
normalised to the brightest cluster (dotted line).

The open squares in Fig.~\ref{fig_1} represent clusters from the
{\it preheating} run in da Silva et al. (2004). This simulation
implemented a simple heating mechanism where the gas was impulsively
heated by 1.5 keV per particle at $z=4$. Despite the good overlap for
luminous clusters, the preheating cluster distribution shows a
somewhat larger dispersion than in CLEF and lower SZ fluxes for less
luminous objects. This is because our feedback model has a relatively
larger effect in smaller systems.
The deviation from self-similarity is therefore larger in the CLEF
case than in the preheating simulation of da Silva et al. (2004), where
$Y^{\rm int}_{200}\propto L_{\rm x,200}^{0.98}$.

\section{Simulation maps}

One important aspect of the CLEF run is that it was designed with
abutting outputs to permit the construction of simulated maps. Direct 
application of the map-making method of da Silva et al. (2000)
to this run allows us to obtain maps of the SZ effect with 5 deg$^2$ of
size, with the signal integrated up to $z=4.6$.
\begin{figure}[t]
\centering
\includegraphics[width=12.49cm]{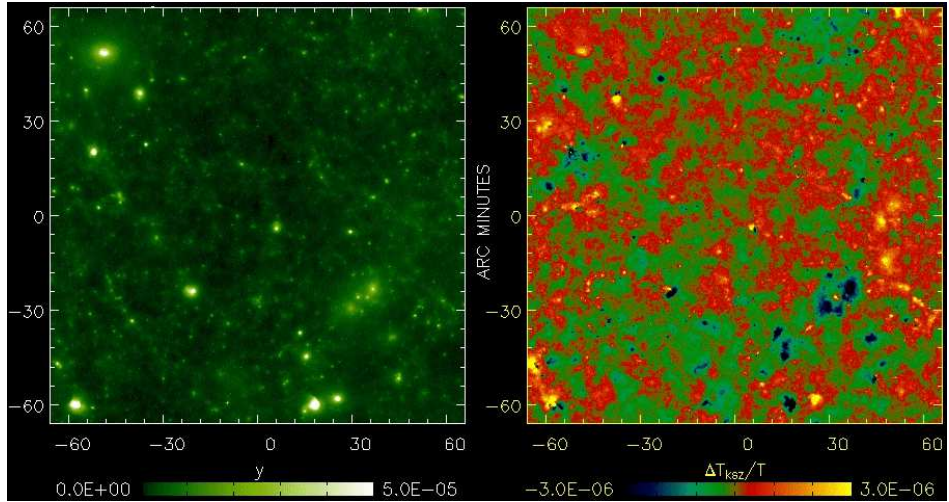}
\caption{Five square degree maps of the thermal $y$-comptonization
parameter (left panel) and kinetic temperature fluctuations (right
panel) of the SZ effect from the CLEF hydrodynamics simulation. Maps
show the same sky realisation with a resolution of 5\arcsec .
}
\label{fig_2}
\end{figure}
This is already a major achievement when compared with previous works
(where map sizes are typically one square degree in size), which
demonstrates one of the major advantages of having a much (typically
eight times) larger simulation volume in CLEF. Maps constructed from
CLEF include more massive structures and large-scale power than in
previous works. Moreover the effect of non-gravitational heating was
also lacking, or only crudely estimated, in most of these simulations.

In Fig.~\ref{fig_2} we present the first thermal (left panel) and
kinetic (right panel) SZ maps from the CLEF simulation with 5 deg$^2$
of size and (raw) resolution of 5 arc-seconds. The mapped quantities
are the $y$-comptonization parameter, for the thermal effect, and the
temperature fluctuation $\Delta T_{\rm ksz}/T$, for the kinetic
effect. These were computed in the usual way --- see da Silva et al.
(2001). The mean $y$-distortion, averaged over lines of sight, gives
an important measure of the global state of the gas in the
Universe. In the CLEF run we find $y_{\rm mean}=3.6\times 10^{-6}$,
obtained by averaging over 60 square degrees of area (12 sky
realisations, each with 5 deg$^2$). This comfortably satisfies 
{\it COBE--FIRAS} constraints on $y_ {\rm mean}$ (Fixsen et al.~1996).
 
Our primary objectives with the CLEF hydrodynamics run are to create
realistic simulations of galaxy clusters and large maps of the SZ
effect for dedicated instrument studies. Templates of simulated
cluster maps can also be created from CLEF to improve present all-sky
simulations of the SZ effect. All these aspects are important for a
better understanding of galaxy cluster observations and the
preparation of future CMB/SZ missions such as Planck, AMiBA and SPT.


\end{document}